\documentclass{aastex631}
\shorttitle{ENERGETIC ELECTRONS AS A DRIVER OF SUNQUAKES}
\shortauthors{Wu et al.}

\begin{document}

\title{Highly Energetic Electrons Accelerated in Strong Solar Flares as a Preferred Driver of Sunquakes}

\author[0000-0002-0786-7307]{H.~Wu}
\affiliation{School of Astronomy and Space Science, Nanjing University, Nanjing 210023, People's Republic of China}

\author[0000-0001-9856-2770]{Y.~Dai}
\affiliation{School of Astronomy and Space Science, Nanjing University, Nanjing 210023, People's Republic of China}
\affiliation{Key Laboratory of Modern Astronomy and Astrophysics (Nanjing University), Ministry of Education, Nanjing 210023, People's Republic of China}

\author[0000-0002-4978-4972]{M.~D.~Ding}
\affiliation{School of Astronomy and Space Science, Nanjing University, Nanjing 210023, People's Republic of China}
\affiliation{Key Laboratory of Modern Astronomy and Astrophysics (Nanjing University), Ministry of Education, Nanjing 210023, People's Republic of China}

\correspondingauthor{Y.~Dai}
\email{ydai@nju.edu.cn}

\begin{abstract}

Sunquakes are enhanced seismic waves excited in some energetic solar flares. Up to now, their origin has still been controversial. In this Letter, we select and study 20 strong flares in Solar Cycle 24, whose impulse phase is fully captured by the \emph{Reuven Ramaty High Energy Solar Spectroscopic Imager} (\emph{RHESSI}). For 11 out of 12 sunquake-active flares in our sample, the hard X-ray (HXR) emission shows a good temporal and spatial correlation with the white-light (WL) enhancement and the sunquake. Spectral analysis also reveals a harder photon spectrum that extends to several hundred keV, implying a considerable population of flare-accelerated nonthermal electrons at high energies. Quantitatively, the total energy of electrons above 300 keV in sunquake-active flares is systematically different from that in sunquake-quiet flares, while the difference is marginal for electrons above 50 keV. All these facts support highly energetic electrons as a preferred driver of the sunquakes. Such an electron-driven scenario can be reasonably accommodated in the framework of a recently proposed selection rule for sunquake generation. For the remaining one event, the sunquake epicenter is cospatial with a magnetic imprint, i.e., a permanent change of magnetic field on the photosphere. Quantitative calculation shows that the flare-induced downward Lorentz force can do enough work to power the sunquake, acting as a viable sunquake driver for this specific event.
\end{abstract}


\section{Introduction} \label{sec:intro}
It is believed that solar flares are a result of rapid release of free magnetic energy stored in the solar corona. Through magnetic reconnection,  the magnetic energy is converted to a variety of forms, which are transported both upward to the interplanetary space and downward to the solar lower atmosphere. In some energetic flares, the flare-powered perturbations can reach the dense photosphere to enhance the local helioseismic waves, which further penetrate through the solar interior and get reflected back to the photosphere, termed as ``sunquakes"  \citep{Wolff1972ApJ}. The first sunquake observation was reported in \cite{Kosovichev1998Natur}, where the wave signature is manifested as circular ``ripples" in Dopplergrams. Since then, more and more such sunquake events have been discovered \citep[e.g.,][]{Donea1999ApJ, Kosovichev2006SoPh, Zharkov2011ApJ}.

Up to now, the origin of sunquakes has still been controversial. Several categories of driving mechanisms have been proposed. The first category assumes flare-accelerated particles as the driver of sunquakes. The sunquakes are excited either by direct impact of the energetic particles on the photosphere  \citep{Kosovichev1998Natur, Kosovichev2007ApJ, Zharkova2007ApJ, Kosovichev2006SoPh, Zharkova2008SoPh}, or due to pressure pulse from the heated chromosphere by thick-target bremsstrahlung of the nonthermal electrons \citep{Donea2006ASPC,Lindsey2008SoPh}. This scenario is analogous to the mechanism for white-light flares (WLFs) of type I \citep{Hudson1972SoPh, Chen2005ApJ, Chen2006ApJ}, and is supported by a good correlation between the sunquake source, white-light (WL) enhancement, and hard X-ray (HXR) emission revealed in many observations \citep{Buitrago-Casas2015SoPh}. In another category, it is assumed that a downward Lorentz force resulting from abrupt and permanent changes of the photospheric magnetic field, which often occur in strong flares \citep{Sudol2005ApJ, Petrie2010ApJ, Fisher2012SoPh, Sun2017ApJ}, can act as a sunquake driver \citep{Hudson2008ASPC, Fisher2012SoPh}.

It has been shown that sunquakes tend to occur in strong flares \citep{Sharykin2020ApJ}. Nevertheless, only a fraction of strong flares can produce a sunquake. Based on a statistical study of major flares in Solar Cycle 24 observed by the \emph{Solar Dynamics Observatory} \citep[\emph{SDO};][]{Pesnell12SoPh} mission, \citet[][hereafter CZ21]{Chen2021ApJ} proposed a selection rule for sunquake generation: a sunquake is more likely to occur when the photosphere shows a net downward oscillatory velocity. In such a case, the photospheric oscillation can be amplified by the in-phase flare-excited impulse, facilitating the generation of a sunquake. Otherwise, the background oscillation should be weakened instead. This may explain the relative rarity of sunquakes in real observations.

The selection role proposed by CZ21 provides a promising explanation for the occurrence rate of sunquakes. However, the detailed mechanisms for sunquake generation are still poorly understood without resorting to other complementary observations. In this Letter, we further include HXR imaging and spectroscopic data to the sample sunquakes analyzed by CZ21, mainly focusing on the possible role of flare-accelerated electrons in producing the sunquakes.

\section{Instruments and Dataset} \label{sec:dat}

The data used in this study mainly come from the Helioseismic and Magnetic Imager \citep[HMI;][]{Schou2012SoPh} on board \emph{SDO} and the \emph{Reuven Ramaty High Energy Solar Spectroscopic Imager}  \citep[\emph{RHESSI};][]{Lin2002SoPh}. HMI measures full-disk Stokes profiles of the \ion{Fe}{1} 6173 {\AA} line with a pixel size of 0.5\arcsec \ and cadence of 45 s, from which data products such as the continuum intensity ($I_c$), Doppler velocity, and vector magnetic field of the photosphere can be derived.  \emph{RHESSI} is designed for imaging and spectroscopic observations of the Sun in X-rays and $\gamma$-rays. Using a rotation modulation of nine detectors with a 4s period, the spacecraft achieves a spatial resolution as high as 2.3\arcsec \ and spectral solution of 1--10 keV over an energy range from 3 keV to 17 MeV.

We start from the sample of events originally compiled in CZ21, which includes the strongest 60 flares in Solar Cycle 24 that occur within 75$\degr$ in longitude. This yields a lower limit of M6.3 in \emph{GOES} soft X-ray (SXR) class for the candidate flares.  As revealed in the HMI $I_c$ images, all of the flares are strong enough to exhibit a distinguishable WL emission enhancement, indicative of WLFs with the potential to produce sunquakes. Furthermore, the flare locations not too close to the limb ensure that the parameters of the possible sunquakes can be credibly derived from the reconstructed HMI egression power maps.

To investigate the possible role of flare-accelerated electrons in generating sunquakes, we focus on flares whose impulsive phase is fully captured by \emph{RHESSI}.  We need to apply such an additional selection criterion since \emph{RHESSI} observations are routinely affected by orbit night and/or other gaps. Doing so reduces the original sample to 20 flare events, of which 12 flares are in association with at least one sunquake, while the remaining 8 ones are seismically quiet. If there are more than one sunquake events in a sunquake-active flare, we consider the most energetic one, which is usually significantly stronger than the others.  The general information of the flares under study, as well as their characteristics to be quantified in the following analysis, are listed in Table 1. Here the sunquake information is adopted from CZ21. We note that all but one (associated with the 2011 August 9 X6.9 flare, No. 4) sunquakes in our list show a net downward oscillatory velocity (in either the 3--5 mHz frequency band or the 5-7 mHz one, or both) during the flare impulsive phase.

\section{Analysis and Results} \label{sec:res}
\begin{figure}
\includegraphics[width=0.9\textwidth]{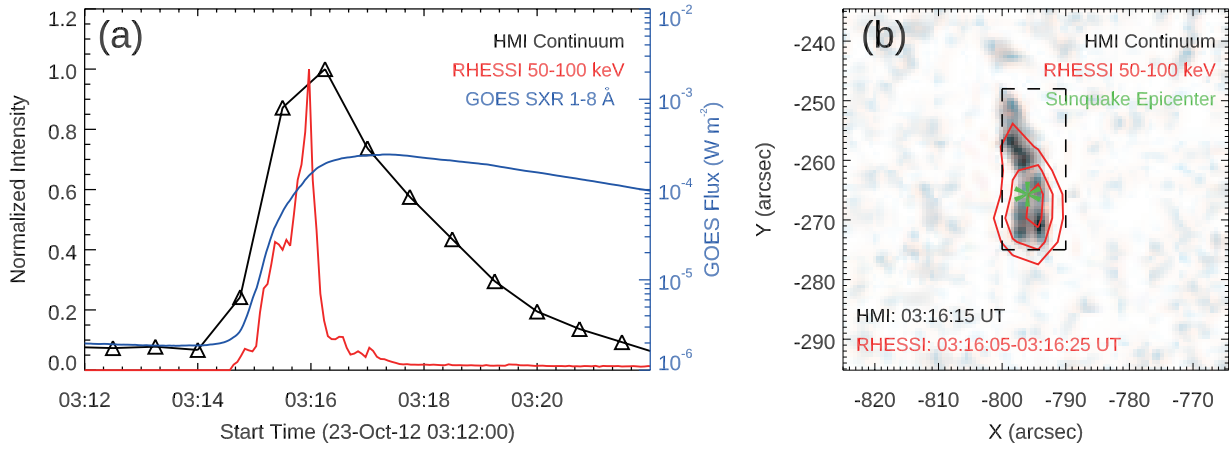}
\caption{WL and X-ray observations of the 2012 October 23 X1.8 flare. (a) time profiles of the HMI continuum intensity around 6173 {\AA} (black), \emph{RHESSI} HXR count rate at 50--100 keV (red), and \emph{GOES} SXR flux in 1--8 {\AA} (blue). (b) the base-difference HMI continuum map at the continuum peak time in an inverse scale, where the dashed box encloses the main flaring region used for continuum calculation. Overplotted on the map is a simultaneous \emph{RHESSI} 50--100 keV image reconstructed using the Pixon algorithm, with contour levels corresponding to 30\%, 60\%, and 90\% of the maximum intensity, respectively.  For this sunquake-active flare, the location of the sunquake epicenter is also marked out with an asterisk sign.}
\label{fig:ichxr}
\end{figure}

Figure \ref{fig:ichxr} depicts the WL and X-ray observations of a typical sunquake-active flare that occurred on 2012 October 23 (No. 7) in NOAA active region 11598. The event has been extensively studied in the literature \citep[e.g.,][]{Yang2015ApJ,Sharykin2017ApJ,Watanabe2020ApJ}, and was also selected as a typical example presented in CZ21.  According to the \emph{GOES} 1--8 {\AA} light curve (blue) plotted in Figure \ref{fig:ichxr}(a), the SXR flare starts at 03:14 UT, promptly rises to its peak at 03:17 UT, and ends at 03:21 UT, registered as an X1.8-class flare. The HXR emission of the flare, as revealed from the \emph{RHESSI} 50--100 keV count rate (red line in Figure \ref{fig:ichxr}(a)), exhibits an even more impulsive increase and peaks at around 03:16 UT, slightly earlier than the SXR emission, which implies that the ``Neupert effect" \citep{Neupert1968ApJL} applies to this flare.  It is also seen that the flare WL emission, which is proxied by the HMI continuum intensity (black line with triangle symbols in Figure \ref{fig:ichxr}(a)) summed over the main flaring region (dashed box in Figure \ref{fig:ichxr}(b)), shows a nearly synchronous enhancement with the HXR emission before reaching its maximum at 03:16:15~UT. After then, the WL emission turns to a relatively gradual decay in comparison with the precipitous drop in HXR emission.

As shown in Figure \ref{fig:ichxr}(b), the WL enhancement at the peak is predominately manifested as two quasi-parallel flare ribbons. Here, for clarity of viewing, we subtract a pre-flare image from the image at the flaring time to highlight the WL enhancement, and plot the base-difference map in an inverse color scale where dark features indicate brightening. When overplotting a simultaneous \emph{RHESSI} image at 50--100 keV (red contours) on the HMI WL map, it is seen that the HXR source well covers the WL ribbons, although the former seems more diffuse. According to \citet{Yang2015ApJ}, the WL ribbons correspond to the western segments of a pair of inner/outer circular ribbons that outline the base of a fan-spine topology, while the HXR source is located around the south footpoint of a magnetic flux rope embedded under the fan dome. The close temporal and spatial correlation between the WL and HXR emissions indicates that this event belongs to a type I WL flare, in which the WL emission originates from the layers heated by a direct electron bombardment and/or the following backwarming effect \citep{Hudson1972SoPh, Chen2005ApJ, Chen2006ApJ,Hao2012AA}. 

For this sunquake-active flare, we also mark out the location of the sunquake epicenter (green asterisk in Figure \ref{fig:ichxr}(b)). As CZ21 have verified a tight correlation between the WL enhancement and sunquake excitation, our complementary HXR observations strongly suggest the same electron-driven scenario for the sunquake generation as that for the WL enhancement in this flare \citep{Sharykin2017ApJ,Watanabe2020ApJ}. By checking other sunquake events, we find that all but one (the 2011 August 9 X6.9 flare, No. 4) of the sunquakes in our list show a good correlation with the HXR emission both temporally and spatially, which further corroborates nonthermal electrons as a preferred driver of the sunquakes. 

To further quantify the energetics of flare-accelerated electrons, we fit the \emph{RHESSI} spectra during the whole flare impulsive phase (listed in Table \ref{tab:summary}) using the Object Spectral Executive (OSPEX) package. First, we divide the impulsive phase into several time intervals, each of which has a duration of 20 s. Then we use a thick-target bremsstrahlung model (\texttt{thick2}), which assumes a broken power-law distribution of the flare-accelerated nonthermal electrons, plus a single-temperature thermal model (\texttt{vth}) to perform the spectral fitting for each individual interval. Since we are only concerned with nonthermal properties, the thermal component is introduced just to better constrain the low-energy cutoff ($E_c$) of the nonthermal electrons. Therefore, the lower limit of the energy range for fitting is fixed at 10~keV to exclude the Fe/Ni emission lines at $\sim$6.7~keV,  which permits a simplification of the thermal component fitting by only varying the temperature and emission measure while keeping the elemental abundance unchanged. On the other end, the upper limit is determined such that the photon flux at that energy starts to drop below the background level.  

\begin{figure}
\includegraphics[width=1\textwidth]{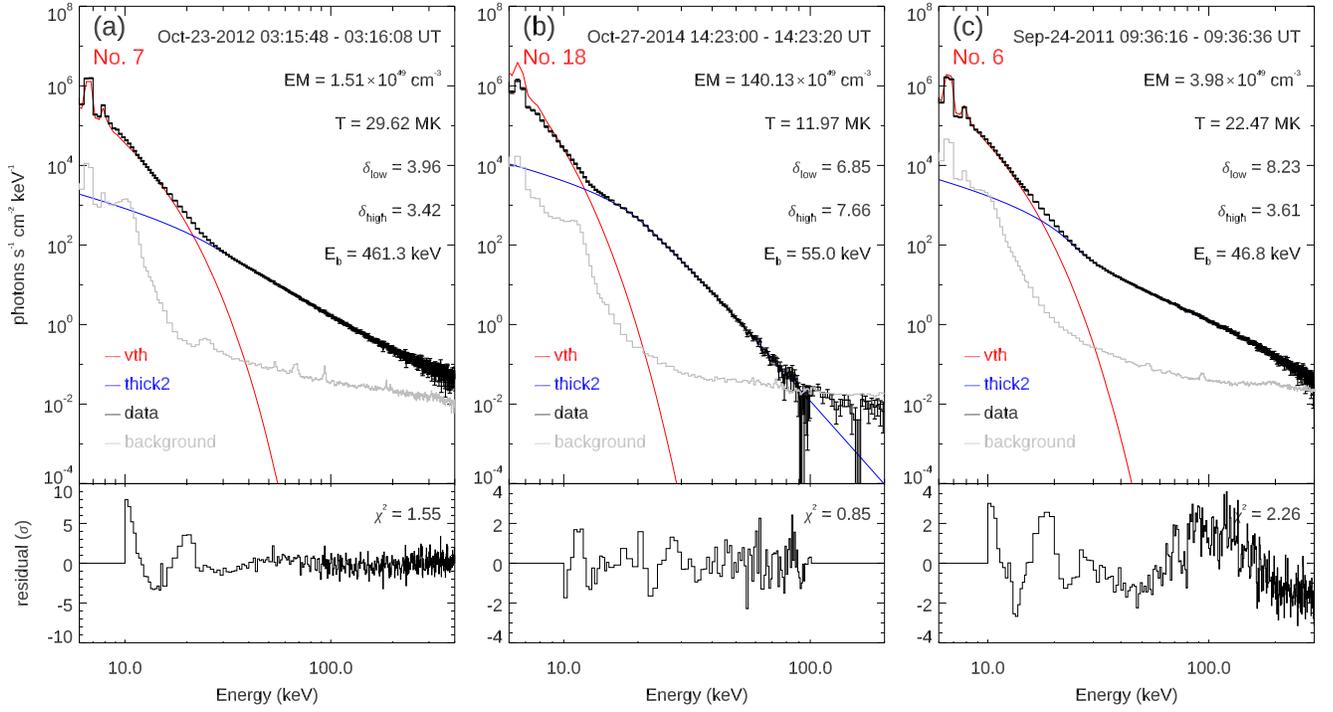}
\caption{Fitting results for the \emph{RHESSI} spectra taken around the HXR peak in three flares. In each panel, the event number in Table \ref{tab:summary} is labeled in the upper left, the black and grey lines in histogram mode denote the background-subtracted photon flux and the background, respectively, while the colored curves represent different components of the modeled spectrum based on the best-fit parameters. In addition, the residual between the modeled and observed spectra is plotted in the bottom part of each panel.}
 \label{fig:spectra}
\end{figure}

Figure \ref{fig:spectra}(a) shows the \emph{RHESSI} spectrum around the HXR peak of the 2012 October 23 flare, as well as the spectral fitting results. It is seen that the photon flux at 30 keV is as high as 68.9 photon~s$^{-1}$~cm$^{-2}$~keV$^{-1}$, among the typical values observed in WLFs \citep{Kuhar2016ApJ,Hao2018NC}. More importantly, the flux keeps above the background level until 400 keV, indicative of a significant fraction of electrons accelerated to very high energies. We note that this is a common spectral feature for the sunquake-active flares. The spectral fitting reveals power-law indices of 3.96 and 3.42 for the nonthermal electrons below and above a break energy of 461 keV, respectively, reflecting a hardening of the spectra toward higher energies. 

For comparison, we also present  in Figures \ref{fig:spectra}(b) and (c) the spectra of the other two flares that are of similar \emph{GOES} classes but without sunquakes.  For these sunquake-quiet flares, the photon flux at 30 keV is comparable to that for the sunquake-active events. Toward higher energies, however, the HXR spectrum shows diverse variations either becoming very soft such that the flux quickly drops below the background (the 2014 October 27 X2.0 flare, No. 18), or still behaving like that of the sunquake-active events (the 2011 September 24 X1.9 flare, No. 6). Obviously, the diverse spectral patterns imply that the population of high energy electrons in sunquake-quiet flares can be distinctly different from case to case.

Based on the spectral fitting, we evaluate the total energy of nonthermal electrons using the integral 
\begin{equation}
E=\int\!\!\!\int \varepsilon F(\varepsilon,t)\, d\varepsilon dt,
\end{equation}
where $\varepsilon$ is the electron energy and $F(\varepsilon,t)$ the fitted electron spectrum. The integration with respect to time is done over the entire flare impulsive phase. As to the energy range for integration, we adopt fixed lower limits regardless of the variable low-energy cutoffs derived from actual flares.  Here we calculate the total energies of the electrons above 50 keV ($E_{50}$) and that above 300~keV ($E_{300}$), which characterize the energetics of mildly and highly energetic electrons, respectively. 

\begin{figure}
    \includegraphics[width=1\textwidth]{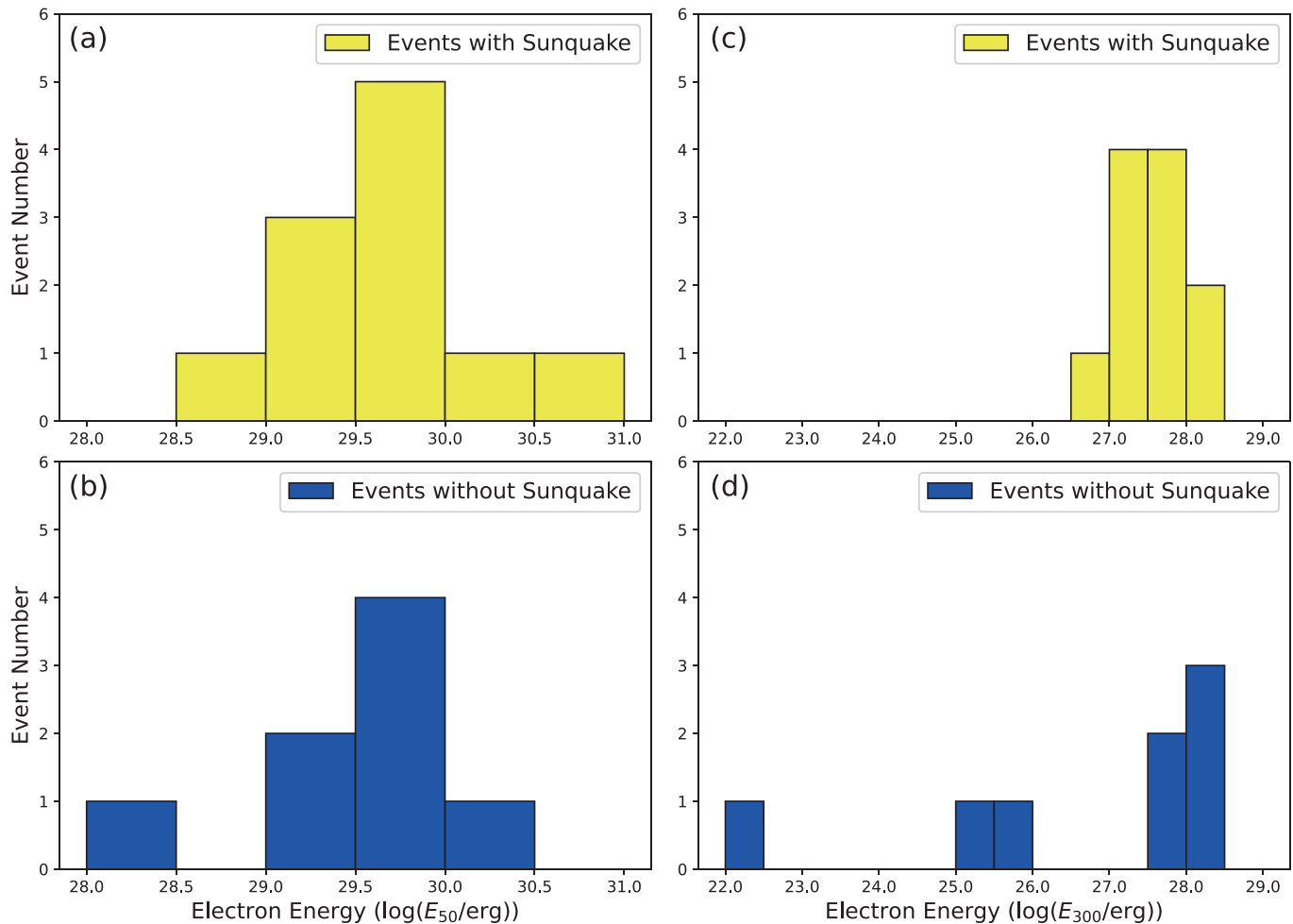}
    \caption{Histograms of the total energy of nonthermal electrons for the flare events with (upper) and without (lower) sunquakes. The left  panels are for the distributions of $E_{50}$ while the right for $E_{300}$. Note that a bin size of 0.5 dex is adopted for the histogram plotting.}
    \label{fig:histo}
\end{figure}

Figure \ref{fig:histo} displays the histograms of $E_{50}$ (left) and $E_{300}$ (right) for the flares with (upper) and without (lower) sunquakes, respectively. Note that we exclude the 2011 August  9 sunquake-active flare in which the sunquake originates in a different place from that for the nonthermal electrons.  It is found that the distribution of $E_{50}$ for sunquake-active flares shows no significant difference from that for sunquake-quiet flares; both distributions span over a similar energy range and peak at 10$^{29.5}$--10$^{30}$~erg (Figures \ref{fig:histo}(a) and (b)). Nevertheless, a systematic difference is seen in the distribution of $E_{300}$. The $E_{300}$ value for the flares with sunquakes varies in a relatively narrow range, and is dominantly restricted to a magnitude of  $10^{27}$--$10^{28}$~erg (Figure \ref{fig:histo}(c)), which is comparable to the estimated energy of sunquakes reported in previous studies  \citep{Donea2006SoPh,Chen2021ApJ}. By contrast, the value of $E_{300}$ for the sunquake-quiet flares seems more scattered, which is either comparable to that for the sunquake-active flares, or several orders of magnitude lower (Figure \ref{fig:histo}(d)). Such a bimodal distribution can be expected from the spectral fitting for the sunquake-quiet flares shown in Figure \ref{fig:spectra}.

{We} also calculate the corresponding electron power, which is obtained by dividing the total electron energy by the duration of impulsive phase. As shown in Table \ref{tab:summary}, the length of impulsive phase just varies in a narrow range of 60--120 s from event to event.  It is found that the distributions of the electron power (not shown here) are nearly the same as those shown in Figure \ref{fig:histo}.

The above statistical result implies that the generation of the sunquakes is more relevant to highly energetic electrons rather than electrons at moderate energies. However, the latter is more likely to be responsible for the enhancement of WL emission. Furthermore, the electron-driven scenario for sunquakes can be reasonably accommodated in the frame of the selection rule proposed by CZ21. In addition to being in phase with the background oscillation, the downward electron beam should contain enough highly accelerated electrons in order to efficiently perturb the photosphere and deep layers to produce a sunquake. As for the sunquake-quiet flares, however, either the electron-driven impulse is too weak (e.g., the 2014 October 27 X2.0 flare shown in Figure \ref{fig:spectra}(b)), or the impulse is out of phase with the background oscillation (e.g., the 2011 September 24 X1.9 flare flare shown in Figure \ref{fig:spectra}(c)), thus unable to generate a sunquake. This is also the reason why the distribution of $E_{300}$ is more scattered for the flares without sunquakes.

Among all the sunquake-active events, the 2011 August 9 flare is an exception in that its sunquake epicenter is spatially offset with the HXR source, which requires an alternative explanation for the sunquake generation. Previous observations have shown that some major solar flares can leave magnetic imprints (MIs) on the photosphere, which are manifested as rapid and irreversible changes of the photospheric magnetic field \citep{Lu2019ApJ}. During this process, the photospheric magnetic field becomes more horizontal, producing a downward Lorentz force on the photosphere that possibly drives a sunquake \citep{Hudson2008ASPC}. In the following, we test the possibility of flare-induced Lorentz force as the sunquake driver for this specific event.

To depict the MIs accurately, we use Space-weather HMI Active Region Patch \citep[SHARP;][]{Bobra2014SoPh} products, whose data pipeline includes a remapping of the magnetic field vector in a cylindrical equal-area (CEA) projection. The three components of the SHARP magnetic field vector are represented by $B_r$ (radial), $B_p$ (southward), and $B_t$ (westward), respectively, from which the magnitude of the horizontal magnetic field is derived as $B_h = \sqrt{B_p^{2} + B_t^{2}}$. Since the flare-induced magnetic field change is mainly reflected in an increase of the horizontal magnetic field, we use regions where $\delta B_h$ exceeds a threshold (e.g., 300 G) to approximate the spatial extent of MIs  \citep[cf.][]{Lu2019ApJ}.

\begin{figure}
    \includegraphics[width=.9\textwidth]{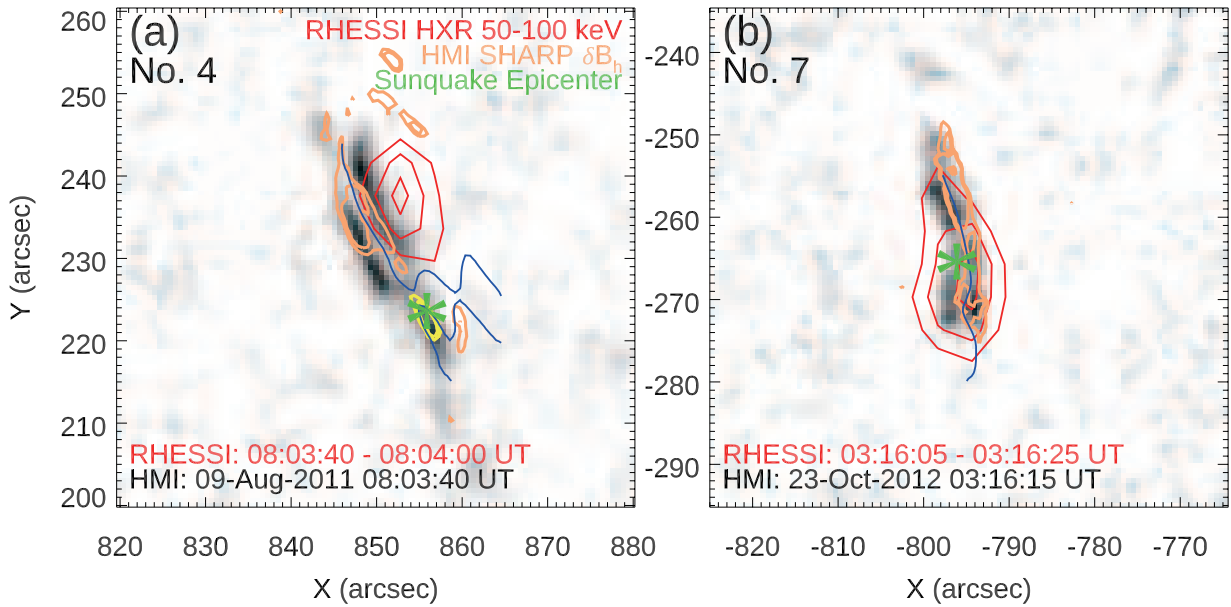}
    \caption{Locations of the MIs (orange plus yellow contours),  HXR source (red contours), and sunquake epicenter (green asterisk) for two flares. In each panel the background image is a corresponding HMI continuum map, with the PIL drawn in blue line. The contours for MI indicate an increase of the horizontal magnetic field at levels of 300 G and 600 G, respectively. Note that the sunquake-related MI in panel (a) is highlighted in yellow contours.}
    \label{fig:sharp}
\end{figure}

We plot in Figure \ref{fig:sharp}(a) the locations of the MIs (orange plus yellow contours),  HXR source (red contours), and sunquake epicenter (green asterisk) for the 2011 August 9 flare, which are overlaid on the corresponding HMI continuum map. As shown in the figure, the MIs appear patch-like, and are located predominately in the vicinity of or over the polarity inversion line (PIL) of SHARP $B_r$, consistent with many previous observations \citep[e.g.,][]{Petrie2012ApJ, Petrie2013SoPh, Wang2012ApJa, Wang2012ApJb, Sun2012ApJ}. The sunquake epicenter lies exactly in a southern MI (distinguished with the other MIs in yellow contours) but distant from the HXR source, which does suggest a Lorentz force-driven origin of the sunquake. 

Compared with other MIs, the sunquake-related MI is located in an isolated region near the far end of the PIL, where the background magnetic field is relatively weaker than that in the AR core. In addition, it appears neither too diffuse nor too compact. These facts may reflect necessary physical conditions for an MI to generate sunquakes. Nevertheless, without other observations of such MI-related sunquakes our argument is not conclusive.

Quantitatively, we use the equation 
\begin{equation}
\delta F = \frac{1}{8 \pi} \int_{A_{ph}}\!(\delta B_r^{2} - \delta B_h^2)\, dA 
\end{equation}
to calculate the Lorentz force $\delta F$ over this sunquake-related MI \citep{Hudson2008ASPC}. When considering an MI area of $A_{ph}=1.3\times10^{17}$~cm$^2$ surrounding the sunquake epicenter if we select a threshold of $\delta B_h=300$~G (enclosed by the outermost yellow contour), the resultant downward Lorentz force on this area is $1.2\times10^{22}$~dyne. By further assuming a displacement of 3~km that the Lorentz force pushes the photosphere downward \citep[cf.][]{Hudson2008ASPC}, we derive a work of  $3.8 \times10^{27}$~erg done by the Lorentz force, which is close to the sunquake energy for this event estimated in CZ21. Compared with the impulsive perturbation by energetic electrons, the MI-induced Lorentz force should act on the photosphere in a much more gentle manner. We note that this sunquake event presents a nearly zero net oscillatory velocity in contrast to the other events. 

Finally we present the corresponding observations of the 2012 October 23 event in Figure \ref{fig:sharp}(b) for comparison. Although the MIs in this flare still gather along the PIL, the sunquake epicenter shows an offset with respect to the MIs in spite of a significant line-shortening due to a close-to-the-limb location of the flare. Instead, the sunquake site should be located in the inner circular flare ribbon.

\section{Discussion and Conclusion} \label{sec:discon}

In this Letter, we make a statistical study on sunquake generation using a sample of 20 strong solar flares that have a full \emph{RHESSI} coverage of the impulsive phase. For 11 out of 12 sunquake-active flares in our sample, the HXR emission shows a good temporal and spatial correlation with the WL enhancement and the sunquake. Spectral analysis also reveals a hard photon spectrum in which the photon flux is well above the background level until several hundred keV, implying a significant population of flare-accelerated nonthermal electrons at high energies. Furthermore, the total energies of electrons above 300 keV in sunquake-active flares are systematically different from those of sunquake-quiet flares, while the difference is marginal for energies above 50 keV. All these facts support highly energetic electrons as a preferred driver of the sunquakes.  Besides the selection rule proposed in CZ21, i.e., the flare-induced impulsive heating should be in phase with a downward background oscillation, a strong electron beam with in particular a significant fraction of energy residing in highly energetic electrons should serve as another necessary condition for the sunquake generation. If either of the two conditions is broken down, a sunquake is not likely to occur.

According to \cite{Neidig1989SoPh}, only electrons above an energy of $\sim$900 keV can penetrate to the photosphere. Nevertheless, in a flaring atmosphere, the ionization, condensation, and evaporation of plasma may mitigate the energy requirement for the electrons to reach such depths \citep{Watanabe2020ApJ}. In this meaning, the electron-driven sunquakes in our sample could be excited by the direct impact of extremely energetic electrons on the photosphere \citep{Kosovichev1998Natur, Kosovichev2007ApJ, Zharkova2007ApJ, Kosovichev2006SoPh, Zharkova2008SoPh}. Nevertheless, it is also possible that the pressure pulse from the heated chromosphere by less energetic electrons plays a part role\citep{Donea2006ASPC,Lindsey2008SoPh}. Without sophisticated radiative hydrodynamic modeling, we do not intend to clarify the quantitative contributions of these mechanisms for the sunquake generation, which should be case-dependent.

There is also an exceptional event (the 2011 August 9 sunquake) in our sample, whose sunquake epicenter is cospatial with an MI instead of the HXR source. We calculate the Lorentz force due to a permanent change of the photospheric magnetic field over this MI, and estimate the work done by the downward Lorentz force. The quantitative analysis shows that the magnetic reconfiguration can provide enough energy to power the sunquake. Therefore, although we suggest highly energetic electrons as a main driver of sunquakes, we do not rule out the role of flare-induced Lorentz force in some specific events \citep{Hudson2008ASPC, Fisher2012SoPh}.

The properties (location and oscillatory velocity) of the electron-driven sunquakes seem different from those of the MI-related sunquake. Actually, we have checked all electron-driven sunquake events in Table \ref{tab:summary}, none of which shows a spatial correspondence with an MI region. Whether it is of physical significance or just a coincidence, we need more observations to address this issue.

This study only covers a sample of 20 events satisfying our selection criteria that the RHESSI era can provide. In order to reach a more conclusive result, more events are required. \emph{RHESSI} has been decommissioned since 2018. Fortunately, we can make use of imaging and spectroscopic observations with the Spectrometer/Telescope for Imaging X-rays (STIX) on board the newly launched Solar Orbiter (SolO) mission \citep{Krucker2020AA} and the Hard X-ray Imager (HXI) on board the upcoming Advanced Space-based Solar Observatory (ASO-S) emission \citep{Zhang2019RAA}. These new observational facilities will help us better understand the origin of sunquakes.

\begin{acknowledgements}
We are grateful to the anonymous referee for his/her insightful comments and suggestions, which led to a significant improvement of the manuscript. This work was supported by National Natural Science Foundation of China under grants 11733003 and 12127901. Y.D. is also sponsored by National Key R\&D Program of China under grants 2019YFA0706601 and 2020YFC2201201. \emph{SDO} is a mission of NASA's Living With a Star (LWS) program.
\end{acknowledgements}


\clearpage

\begin{deluxetable}{clccccchcccc}
\tablenum{1}
\tablecaption{List of the Flares under study and the Sunquake Information  \label{tab:summary}}
\tablewidth{0pt}
\tablehead{
\colhead{No.} & \colhead{Date} & \colhead{\emph{GOES}} & \multicolumn{4}{c}{\emph{RHESSI} HXR Information} &\colhead{} & \multicolumn{4}{c}{HMI Sunquake Information}  \\
\cline{4-7} \cline{9-12}
\colhead{} & \colhead{} & \colhead{Class} & \colhead{Impulsive Phase} & \colhead{Peak\tablenotemark{\scriptsize a}} & \colhead{${E_{50}}$} & \colhead{${E_{300}}$} & \colhead{} &\colhead{Sunquake} & \colhead{Correlation} & \colhead{$v_{35}$\tablenotemark{\scriptsize b}} & \colhead{$v_{57}$\tablenotemark{\scriptsize b}} \\
\colhead{} & \colhead{} & \colhead{} & \colhead{\footnotesize (UT)} & \colhead{\footnotesize (UT)} &
\colhead{\footnotesize ($10^{30} $ erg)} & \colhead{\footnotesize ($10^{27} $ erg)} & \colhead{} & \colhead{\footnotesize (Y/N)} & \colhead{\footnotesize (HXR/MI)} & \colhead{\footnotesize(m s$^{-1}$)} & \colhead{\footnotesize(m s$^{-1}$)}
}
\startdata
1  & 2011 Feb 13 & M6.6 & 17:33:28--17:34:48 & 17:34:18 & 0.1  & 0.02  && N &     &     &      \\
2  & 2011 Feb 15 & X2.2 & 01:54:24--01:56:04 & 01:55:14 & 0.4  & 1.4   && Y & HXR & 27  & 29   \\
3  & 2011 Jul 30 & M9.3 & 02:07:28--02:08:48 & 02:08:18 & 0.2  & 0.3   && Y & HXR & 417 & 337  \\
4  & 2011 Aug 9  & X6.9 & 08:02:40--08:04:20 & 08:03:50 & 3.2  & 8.9   && Y & MI\tablenotemark{\scriptsize c}  & \nodata & -3   \\
5  & 2011 Sep 6  & X2.1 & 22:18:20--22:19:40 & 22:19:10 & 0.8  & 29.3  && Y & HXR & 326 & 596  \\
6  & 2011 Sep 24 & X1.9 & 09:35:16--09:36:56 & 09:36:26 & 0.5  & 22.6  && N &     &     &      \\
7  & 2012 Oct 23 & X1.8 & 03:15:08--03:16:08 & 03:15:58 & 1.1  & 23.1  && Y & HXR & 1082& 950  \\
8  & 2013 May 15 & X1.2 & 01:41:20--01:43:00 & 01:42:10 & 0.4  & 4.7   && N &     &     &      \\
9  & 2013 Oct 25 & X1.7 & 07:58:10--07:59:50 & 07:59:20 & 0.6  & 9.2   && Y & HXR & \nodata & 135  \\
10 & 2013 Oct 25 & X2.1 & 15:00:12--15:01:52 & 15:00:42 & 0.6  & 5.4   && N &     &     &      \\
11 & 2013 Oct 28 & X1.0 & 01:58:48--02:00:28 & 01:59:38 & 0.3  & 10.6  && N &     &     &      \\
12 & 2013 Nov 10 & X1.1 & 05:12:10--05:13:50 & 05:12:40 & 0.2  & 2.7   && Y & HXR & 445 & 508  \\
13 & 2014 Jan 7  & M7.2 & 10:10:48--10:12:28 & 10:11:38 & 0.5  & 5.6   && Y & HXR & 436 & 680  \\
14 & 2014 Mar 29 & X1.0 & 17:46:00--17:47:40 & 17:46:30 & 0.2  & 11.0  && N &     &     &      \\
15 & 2014 Jun 11 & X1.0 & 09:04:20--09:05:40 & 09:04:50 & 0.06 & 5.6   && Y & HXR & \nodata & 1338 \\
16 & 2014 Oct 22 & M8.7 & 01:38:36--01:40:16 & 01:39:26 & 0.3  & 1.0   && Y & HXR & 133 & -1   \\
17 & 2014 Oct 22 & X1.6 & 14:05:00--14:06:40 & 14:06:30 & 3.9  & 1.4   && Y & HXR & 96  & -41  \\
18 & 2014 Oct 27 & X2.0 & 14:21:20--14:23:20 & 14:23:10 & 1.4  & 0.04  && N &     &     &      \\
19 & 2015 Mar 7  & M9.2 & 22:03:40--22:05:00 & 22:04:30 & 0.01 & 1.4e-5&& N &     &     &      \\
20 & 2017 Sep 7  & M7.3 & 10:14:28--10:16:08 & 10:15:38 & 0.4  & 8.8   && Y & HXR & 534 & 344  \\
\enddata
\tablecomments{
\tablenotetext{a}{ Peak time for HXR emission at 50--100 keV.}
\tablenotetext{b}{Oscillatory velocities at 3--5 MHz ($v_{35}$) and 5--7 MHz ($v_{57}$), respectively. The values are adopted from CZ21.}
\tablenotetext{c}{The estimated work done by the MI-induced Lorentz force is $3.8\times10^{27}$~erg.}
}
\end{deluxetable}

\clearpage

\clearpage

\clearpage

\clearpage

\listofchanges
\end{document}